\documentclass[a4paper,11pt]{article}
\pdfoutput=1 % if your are submitting a pdflatex (i.e. if you have
\usepackage{jcappub} % for details on the use of the package, please
\usepackage[T1]{fontenc} %http://tex.stackexchange.com/questions/664/why-should-i-use-usepackaget1fontenc
\usepackage{lmodern}%http://tex.stackexchange.com/questions/7185/how-to-make-the-pdf-output-of-better-print-quality
\usepackage{hyperref}
\usepackage{graphicx}% Include figure files
\usepackage{bm}% bold math
\usepackage{amsmath}
\usepackage{amssymb}
\usepackage{dcolumn}% Align table columns on decimal point
\usepackage{multirow}
\usepackage{color}
\usepackage[dvipsnames]{xcolor}
\usepackage{xspace}

\newlength{\wdo}
\newcommand{\stroke}[1]{{$#1$}%
\settowidth{\wdo}{${#1}$} {\kern-\wdo}%
\partialvartstrokedint}

\makeatletter
\newcommand{\fancysep}{%
  \@afterindentfalse
  {\begin{center}
    \resizebox{0.8\linewidth}{0.4ex}{{%
        \fontsize{20}{24}\usefont{U}{webo}{xl}{n}{4}}}%
  \end{center}}\@afterheading}
\makeatother

{\vskip 1.0em
\framebox[\columnwidth][r]{%
\begin{minipage}[c]{\columnwidth}%
\vspace{-1.0em}%
#1%
\end{minipage}}}
{\vskip 1.0em}

\def\XXint#1#2#3{{\setbox0=\hbox{$#1{#2#3}{\int}$}
     \vcenter{\hbox{$#2#3$}}\kern-.5\wd0}}

\newcommand{\beq}{\begin{equation}}
\newcommand{\eeq}{\end{equation}}
\newcommand{\beqa}{\begin{eqnarray}}
\newcommand{\eeqa}{\end{eqnarray}}

\newcommand{\nunu}{\ensuremath{\nu\nu}\xspace}
\newcommand{\Hnn}{\ensuremath{{H}_{\nu\nu}}\xspace}
\newcommand{\Hnnb}{\ensuremath{{H}_{\nu\nu}^{\text{bulb}}}\xspace}
\newcommand{\Hnnh}{\ensuremath{{H}_{\nu\nu}^{\text{halo}}}\xspace}

\title{Collective neutrino oscillations with the halo effect in single-angle approximation}

\author{Vincenzo Cirigliano,}

\author{Mark Paris,}

\author{Shashank Shalgar}
\affiliation{Theoretical Division, Los Alamos National Laboratory, Los Alamos, NM-87545 USA}

\emailAdd{cirigliano@lanl.gov}
\emailAdd{mparis@lanl.gov}
\emailAdd{shashankshalgar@gmail.com}

\date{\today}

\abstract{
We perform a self-consistent calculation of  collective neutrino oscillations 
including the effect of back  scattered neutrinos (Halo effect) in the `single-angle' approximation, 
within a spherically symmetric supernova model. 
We find that  due to the Halo effect the onset of flavor transformations 
is pushed to smaller radii, by a few kilometers. 
The celebrated phenomenon of the spectral split is found to be robust
under the present inclusion of the Halo effect. 

}
\begin{document}
\maketitle

\section{Introduction}
The extreme astrophysical environment of a
protoneutron star (PNS), formed within a few 10's of seconds after the
collapse of a massive star (in the range $\sim 8 - 50 \, M_\odot$), in
the aftermath of a core-collapse supernova (CCSN), is largely
characterized by the non-equilibrium evolution of its neutrino field.
Indeed, about $99$\% of the gravitational energy released in the
collapse is converted to neutrinos with almost $\sim 10^{58}$ neutrinos
emitted within the first few seconds; this number
is about 20 orders of magnitude greater 
than the number of neutrinos emitted by the sun in one second.
The reach and importance
of neutrino flavor oscillation effects in astrophysical systems is
difficult to overstate, as two familiar examples illustrate.  Neutrino
heating in CCSN, highly sensitive to neutrino  flavor, may be
pivotal in the determination of whether neutrinos are responsible for
the revival of a stalled core-bounce
shock\cite{Bethe:1984ux, Brandt:2010xa, Janka:2006fh}; the synthesis
of $r$-process elements in extreme astrophysical environments is, too,
expected to be sensitive to such oscillations\cite{Duan:2010af}.

Significant progress in the understanding of the
evolution of the neutrino flavor field and the role of neutrino flavor
oscillations, has been recently achieved\cite{Pantaleone:1994ns,
Duan:2005cp,Duan:2006an,Duan:2006jv,Duan:2007mv,Duan:2008za,
Raffelt:2007xt,EstebanPretel:2007bz,EstebanPretel:2008ni,
Raffelt:2008hr, Dasgupta:2010ae, Malkus:2014iqa,Zhu:2016mwa,Wu:2015fga,
Raffelt:2013rqa, Mirizzi:2013wda, Duan:2014gfa,Abbar:2015mca}.
Particularly interesting is the prospect of collective motion in
flavor space, where the phase and frequency of neutrino flavor
oscillations are independent of their
momenta\cite{Duan:2006an,Duan:2006jv,Duan:2007bt}. The effect is
induced by coherent forward scattering, whose interaction strength is
of bilinear form -- the {\em sine qua non} of dense
neutrino environments. It dictates, in high-flux neutrino
environments, a non-perturbative setting that challenges both
conventional analytical and computational approaches.

In spite of these achievements, a detailed description of the neutrino
environment near the PNS in the aftermath of the CCSN remains
incomplete. For example, the radius (or `epoch'), measured from
the center of the PNS, at which flavor oscillations commence is only
known in the  simplest of models. 
This uncertain situation is largely a product of
limitations imposed by both the complicated geometry of the PNS
environment and the nonlinear nature of the non-equilibrium evolution
of the neutrino flavor field, governed by the neutrino quantum kinetic
equations\cite{Sigl:1992fn,Raffelt:1992uj,McKellar:1992ja,
Enqvist:1990ad,Strack:2005ux,Volpe:2013jgr,Volpe:2015rla,Vlasenko:2013fja,
Zhang:2013lka,Cirigliano:2014aoa,Serreau:2014cfa,Blaschke:2016xxt} (QKE), which limits the use of analytical
methods. We must, therefore, resort to numerical techniques. 

Even then, given the complexity of the computational problem that is
faced in the CCSN/PNS environment, limitations due to computational
resource considerations force one to reduce the full problem of the
calculation of the neutrino flavor density matrix, parameterized
generally by four spacetime coordinates and three momentum variables.
The evolution of the neutrino flavor density matrix in this space of
seven independent variables can be significantly reduced by assuming
spherical symmetry.\footnote{We are, for the present work, ignoring
the fact that inhomogeneity and anisotropy
\cite{Duan:2014gfa,Abbar:2015mca,Cirigliano:2017hmk,Mirizzi:2015fva,
Chakraborty:2015tfa} 
is unstable with respect to
seed fluctuations due to the nonlinear nature of the neutrino QKE.}
The density matrix, under this assumption of spherical symmetry, is
then a function only of 
the zenith angle of the momentum $\vec{p}$,  
the energy $E$ (or  $|\vec{p}|$), 
and the distance $r$ from the center of the PNS.  Realistic scenarios at radii
far from the PNS are considerably more complicated than the simple,
spherically symmetric picture allows.  
While the validity of assuming  a simplified geometry  remains to be 
determined\cite{Raffelt:2013rqa, Mirizzi:2013rla,Duan:2014gfa,Abbar:2015mca,Mirizzi:2015fva}, 
it is hoped that   by considering this simplified setup,
insight may be gained into the
nature of neutrino flavor transformation in the complex astrophysical
environment in the vicinity of the PNS.

The pioneering study of Cherry {\em et al.}\cite{Cherry:2012zw}
explored the consequences of perturbing the neutrino ``bulb'' model,
wherein neutrinos are assumed to trace straight-line trajectories for
radii $r > R_\nu$, defining the ``neutrinosphere.''  
They realized that  even a small number  (about 1 part in $10^3$) of neutrinos that 
suffer direction-changing scatterings, residing in a ``halo'' above the PNS, have the potential
to alter the radial neutrino flavor field evolution significantly.
This is due to the PNS geometry,  the form of the forward-scattering neutrino-neutrino
(\nunu) self-interaction,  and the non-linearity of the problem.
The magnitude of contributions to the {\em diagonal} elements of
the \nunu self-interaction
Hamiltonian \Hnn from the bulb $|\Hnnb|$ and halo $|\Hnnh|$
contributions
were
investigated in Ref.\cite{Cherry:2012zw}; phase differences induced by
the medium during neutrino propagation were neglected and
contributions to $|\Hnnh|$ were assumed to arise from neutrinos
scattered by a single direction-changing scattering.  The tantalizing
results of their study suggest the possibility of a significant role
for the halo in determining the evolution of the PNS neutrino flavor
field with an enhancement of $|\Hnnh|/|\Hnnb|$ by a factor of up to 10
at distances 1000--2000 km from the PNS. The neutrino self-interaction
falls off as the forth power of the radius while the collision term,
which is proportional to the matter density falls off approximately
like the third power of the radius, which increases the relative
strength of the Halo Hamiltonian with respect to the bulb Hamiltonian.
Given this linear increase in the ratio of $|\Hnnh|/|\Hnnb|$, we take as
our departure for the current study the question posed by
Ref.\cite{Cherry:2012zw} of whether scattered-halo feedback effects
push flavor oscillations to smaller radii. We find, in fact, that such
a hastening of the flavor transformation is expected but likely too
small to observe.

In the present work, we investigate the effect of going beyond the approximations of Ref.\cite{Cherry:2012zw}. 
Using a  collision term rooted in the QKE~\cite{Blaschke:2016xxt}, 
we compute the effect of neutrino back-scattering, the halo effect, through an iterative, and therefore
self-consistent, approach within the `single-angle' approximation. 
A broader theme we touch upon is the robustness of the spectral
splits seen in the simulations of the neutrino flavor oscillations in 
the interior of a core-collapse supernova. We find that  the 
spectral split is seen in the final electron neutrino spectrum in the
single-angle approximation even upon the inclusion of the 
backscattered neutrinos.

Relaxing the assumption that those neutrinos which back-scatter from
nuclei do not contribute to the neutrino-neutrino self-interaction
increases the computational complexity of the problem relative to that
of the bulb model. 
We have implemented the effect of neutrino back-scattering in
 the single-angle approximation, using a simplified  collision 
term derived from the full expressions in Ref.~\cite{Blaschke:2016xxt}. 
At each radius there is a finite probability of neutrino back-scattering
and we take into account the effect of the back-scattered neutrinos on the 
neutrino flavor evolution. 
While the neutrino-bulb model is an initial value problem, including the 
back-scattered neutrinos transforms it into a boundary value problem. 

In the following section, we review the essential physics of neutrino
flavor evolution in the bulb model. We detail, in Sec.~\ref{sec:model}, 
the present halo model formulation and solution method. Finally, we
discuss the results (Sec.~\ref{sec:nr}) 
of these calculations and conclude in Sec.~\ref{sec:conc} with a summary
of the main results.

\section{Overview of flavor instability} 
\label{sec:ofi}

We review the neutrino flavor oscillations in the context of the bulb
model, the simplest, perhaps, of models that attempt to capture
coherent oscillation effects in the PNS environment.    The bulb model
assumes a spherical and sharp transition at $R_\nu$ from trapped- to
free-neutrinos.  This approximation should    capture gross features
in the complicated environment, as long as the neutrino flavor
instability has an onset at a relatively large radius compared to
$R_\nu$.  The bulb model  takes into account the effects of \nunu
self-interaction but neglects the effect of neutrinos back-scattered
by matter in the PNS environment. This discussion is intended to
provide an overview of the essential physics of neutrino flavor
instability, which remains largely relevant in the present
calculation of the onset of flavor instability.  
We work with two effective neutrino flavors labeled $e$
and $x$ and describe neutrinos by a $2 \times 2$ density  matrix
$\rho_{\alpha \beta}$,  with $\alpha, \beta \in \{e, x \}$. We assume a value of $R_{\nu} = 11$ km for this work.  

\begin{figure}
\begin{center}
\includegraphics[width=0.49\textwidth]{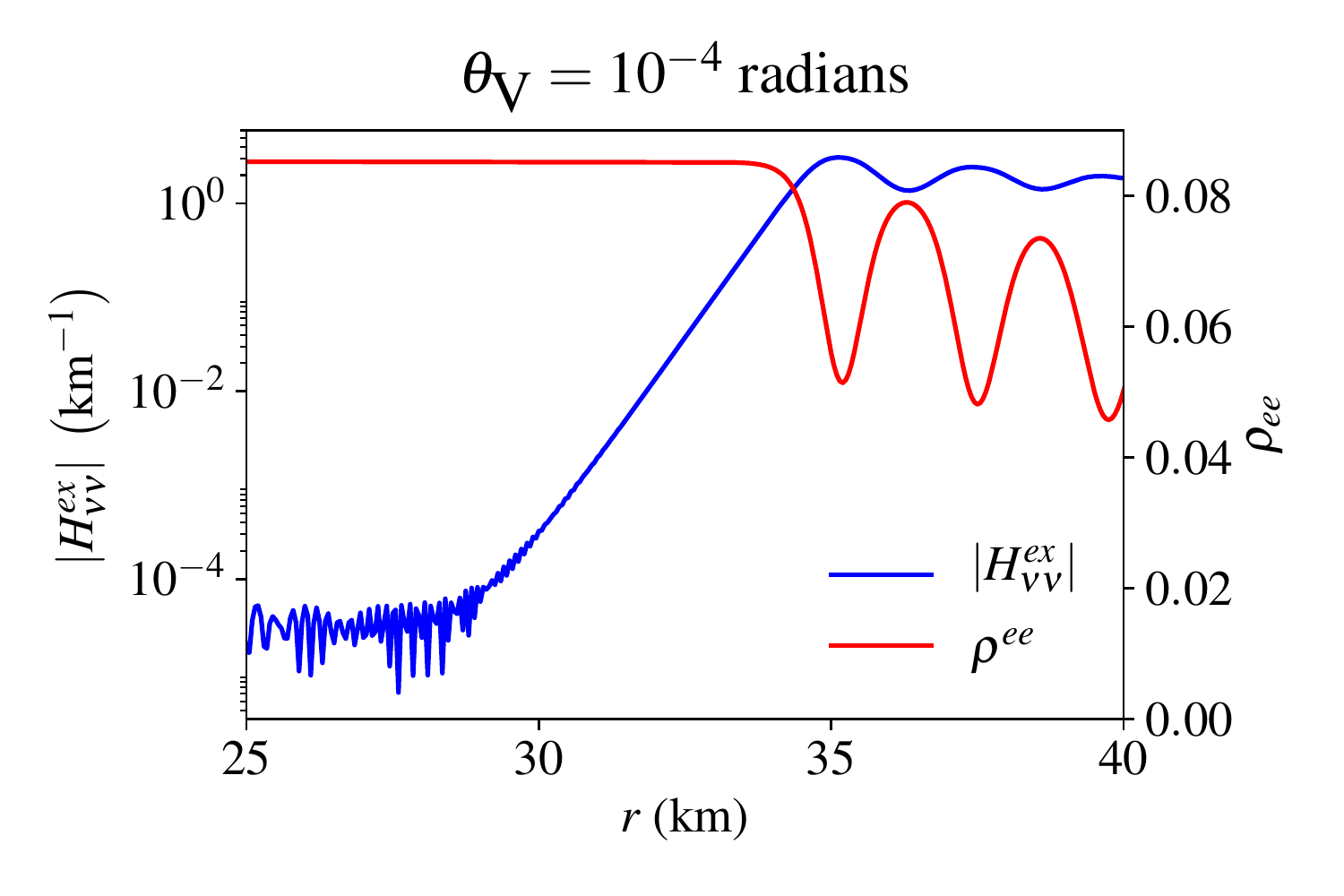}
\includegraphics[width=0.49\textwidth]{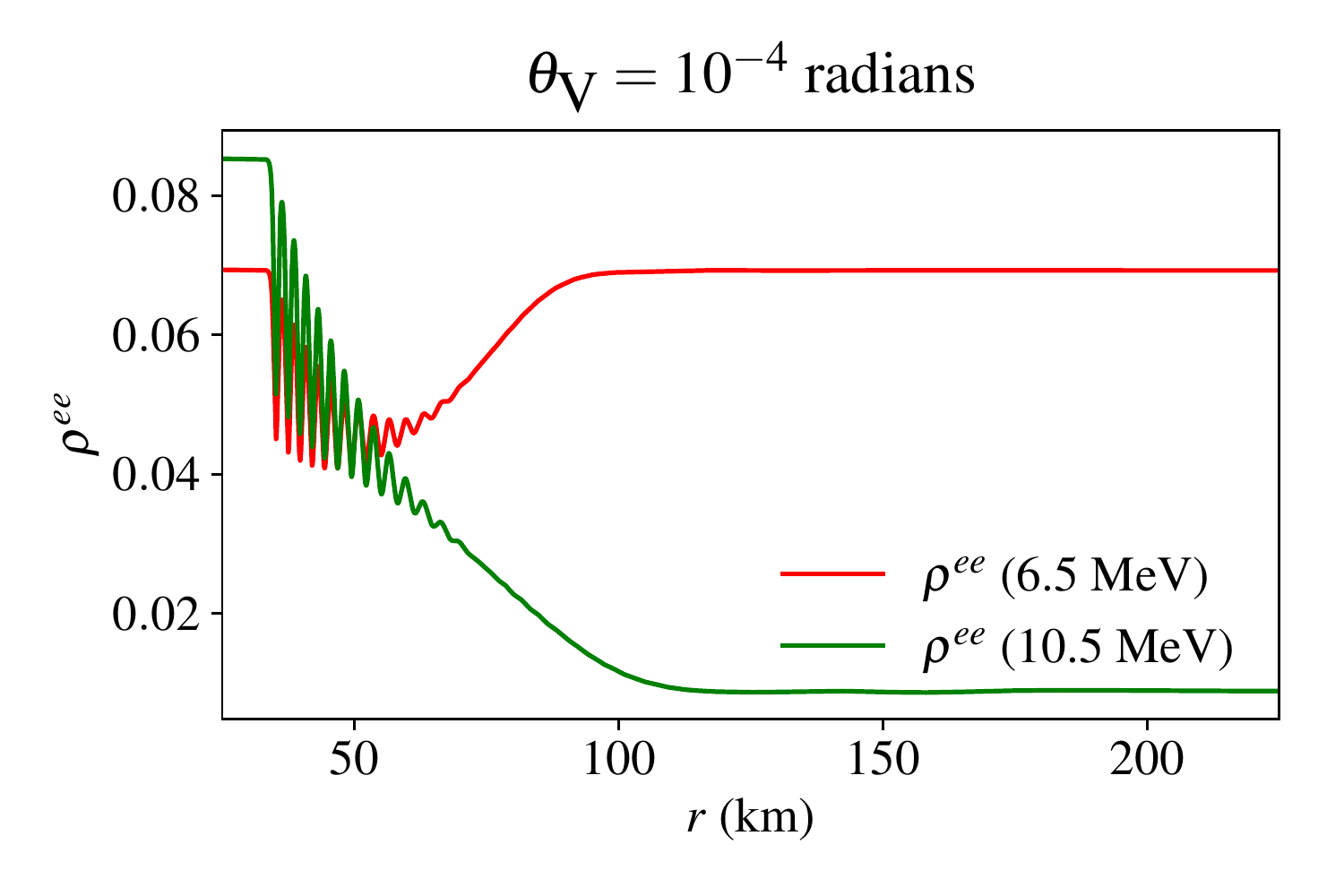}
\caption{Left panel: The logarithm of the off-diagonal term of the \nunu
   Hamiltonian, $\Hnn^{ex}$ (blue curve; color online) and the
   electron neutrino flux, $\rho^{ee}$ (red curve; color
   online) at an energy of 8.5 MeV. The neutrino flavor
   instability, signaled by the variation of the $\Hnn^{e\mu}$ with
   radius, commences at radii $\sim 29$ km. At larger radii, beyond 33
   km, when the off-diagonal term of Hamiltonian attains its maximum
and no longer varies with radius, we observe the onset of neutrino
oscillations. Right panel: We plot electron neutrino flux, 
in arbitrary inverse-energy units,
for two 
representative energies, below and above the split energy. Above the
split energy the electron neutrino flux returns to its initial value. 
}
\label{onset}
\end{center}
\end{figure}

We consider the equations of motion of  the neutrino density matrix 
$\rho$   in the presence of three contributions to the evolution operator $H$: the vacuum
contribution $H_0$, the matter-effect term $H_{\textrm{m}}$, 
and the neutrino self-interaction term $H_{\nu\nu}$. 
The Hamiltonian $H$ can be represented as traceless $2 \times 2$ 
matrix and the amplitude of the 
neutrino flavor transformation is determined by the ratio of the 
off-diagonal element to the diagonal element. 

In the neutrino bulb-model, the strength of the neutrino self-interaction
potential is very large near the proto-neutron star 
($H\sim H_{\nu \nu}$  is diagonal) 
and there is  no flavor instability as a result\footnote{The assumption of
flavor stability at large neutrino flux is not necessarily
valid when the assumptions of neutrino bulb-model are relaxed. These 
assumptions include spherical symmetry and a sharp transition to
free streaming of neutrinos of all flavors at a particular radius. 
}.
The size of the off-diagonal terms of $H$,
in this case, is determined by the vacuum mixing angle as well as the
matter density. As the radial distance increases,  the 
self-interaction decreases until neutrino flavor instability is reached
and the off-diagonal terms of the Hamiltonian grow exponentially.
The onset of instability is controlled by the relative size of 
the vacuum Hamiltonian $H_0$ and 
the diagonal components of $H_{\nu \nu}$.  Similarly, the 
exponential growth is controlled by the diagonal components of 
$H_{\nu \nu}$, as can be seen from a linear stability analysis 
\cite{Banerjee:2011fj}.

Exponential growth of the off-diagonal element of $H$, driven by the
non-linear neutrino self-interaction \Hnn, eventually triggers the
onset of neutrino flavor transformation as seen in Fig.1 at radii 
$\gtrsim 35$ km.
As the distance from the proto-neutron star increases,  the neutrino
flavor reaches a steady state configuration, characterized by a
spectral split ($\nu_e$ and $\nu_x$ spectra above a certain energy
$E_c$ get swapped).  On the right hand side panel of Fig.~\ref{onset} 
we can see this phenomenon in our 
numerical simulations. For a representative energy lower than $E_{c}$
the $\rho_{ee}$ component of the density matrix returns to the initial value,
whereas for a representative energy above $E_{c}$ the flux 
approaches the initial $x$ flux (not shown in the figure).
To summarize, there are two regions, near the
proto-neutron star and away from the proto-neutron star, where the
flavor does not evolve significantly, but for different reasons. 

As we will show later the 
magnitude of modification to the Hamiltonian due to the Halo effect
is of consequence  only in the region of significant
flavor transformation.
In the model we consider, the Halo effect does not lead to 
additional exponential growth due to its 
inclusion. 
We find that the inclusion of the Halo effect does not change the
region of instability or the exponential growth rate of the
off-diagonal term in the neutrino-bulb model with single-angle
approximation. 
 
This occurs because the Halo neutrinos affect the diagonal entries 
of $H_{\nu \nu}$, which control both the onset of instabilities  and their 
growth rate,  only at the percent level for radii of interest.
However, we find that
at small radii  the off-diagonal entries of $H_{\nu \nu}$ are significantly
modified due to the Halo effect.  This in turns can 
change the radius at which the off-diagonal terms of the Hamiltonian
become comparable to the diagonal terms and start to affect the
neutrino flavor content. We demonstrate this effect using numerical
simulations of the equations of motion given in the following
section. 

%**********************************************************************
\section{Model set-up and formulation}
\label{sec:model}
%**********************************************************************
In order to study the Halo effect we consider a simple spherically symmetric model. The neutrino sphere, of radius $R_{\nu}$, is assumed to emit neutrinos with only one emission angle. At each point during its propagation there is a finite probability for the neutrino to scatter backwards along the same path, but in the opposite direction. For simplicity we work in two flavor approximation with a pinched thermal spectrum. 

Our objective is to find a steady state solution for this problem, where the outgoing neutrinos experience self-interaction due to other outgoing neutrino and the back-scattered neutrinos.  
The reflected neutrinos also experience potential due to other reflected neutrino and outgoing neutrinos. In this section, we describe the equations of the motion we use for the system, 
namely the appropriate QKEs.

As discussed earlier, assuming spherical symmetry the  $n_f \times n_f$  density matrices  ($n_f$ is the number of flavors)  depend on  
the neutrino energy $E$,  the angle  $\vartheta$  of the neutrino  momentum  with respect to the radial direction, and the radial coordinate $r$, i.e. $ \rho = \rho (E, \cos \vartheta, r)$.  
In the bulb model, which ignores direction-changing scattering, 
 at a given radius $r$ the angle  $\vartheta$  is  related to the emission angle $\vartheta_{E}$  by
\begin{eqnarray}
\cos\vartheta = \sqrt{1-\frac{R_{\nu}^{2}}{r^{2}}\left(1-\cos^{2}\vartheta_{E}\right)}.
\label{varthetadef}
\end{eqnarray}
In the single-angle approximation we use in this paper, only one emission angle is considered, which we call $\vartheta_{0}$, 
and therefore at a given radius $r$ only a single angle $\vartheta (r, \vartheta_0)$ is allowed, defining a cone around the radial direction. 
We {\it define} the single-angle approximation in presence of 
direction-changing collisions by  enforcing that the scattered neutrinos at a given radius $r$  are restricted to the cone defined by $\vartheta (r, \vartheta_0)$. 
A neutrino  scattered in the forward hemisphere  is characterized by  $\vartheta (r, \vartheta_0)$, while a neutrino scattered in the 
backward hemisphere is characterized by $\pi -\vartheta (r, \vartheta_0)$.  
We use the following simplified notation: 
$\rho (E,  \cos \vartheta, r) \equiv   \rho^{\uparrow} (E,r)$  and 
$\bar{\rho} (E, \cos \vartheta, r) \equiv \bar{\rho}^{\uparrow} (E,r)$ 
 correspond to outgoing neutrinos   and anti-neutrinos,
 while  
 $\rho (E, - \cos \vartheta, r) \equiv   \rho^{\downarrow} (E,r)$  and 
$\bar{\rho} (E, - \cos \vartheta, r) \equiv \bar{\rho}^{\downarrow} (E,r)$ 
correspond to   back-scattered  (anti)neutrinos.

The QKEs governing  the evolution of these  density matrices are given by,
\begin{subequations}
\label{eq:qkes}
\begin{eqnarray}
\label{eom1}
\frac{\partial \rho^{\uparrow}(E,r)}{\partial r} &=& - \frac{i}{\cos\vartheta} [H^{\uparrow},\rho^{\uparrow}(E,r)]  %\nonumber\\
- \mathcal{C}^{\textrm{loss}\uparrow}\rho^{\uparrow}(E,r)+ \mathcal{C}^{\textrm{gain}\uparrow}\rho^{\downarrow}(E,r)\\
\label{eom2}
\frac{\partial \bar{\rho}^{\uparrow}(E,r)}{\partial r} &=& - \frac{i}{\cos\vartheta} [\bar{H}^{\uparrow},\bar{\rho}^{\uparrow}(E,r)]%\nonumber\\
- \mathcal{C}^{\textrm{loss}\uparrow}\bar{\rho}^{\uparrow}(E,r)+\mathcal{C}^{\textrm{gain}\uparrow}\bar{\rho}^{\downarrow}(E,r)\\
\label{eom3}
\frac{\partial \rho^{\downarrow}(E,r)}{\partial r} &=& - \frac{i}{\cos\vartheta} [H^{\downarrow},\rho^{\downarrow}(E,r)%]\nonumber\\
- \mathcal{C}^{\textrm{loss}\downarrow}\rho^{\downarrow}(E,r)+\mathcal{C}^{\textrm{gain}\downarrow}\rho^{\uparrow}(E,r)\\
\label{eom4}
\frac{\partial \bar{\rho}^{\downarrow}(E,r)}{\partial r} &=&- \frac{i}{\cos\vartheta} [\bar{H}^{\downarrow},\bar{\rho}^{\downarrow}(E,r)]%\nonumber\\
- \mathcal{C}^{\textrm{loss}\downarrow}\bar{\rho}^{\downarrow}(E,r)+\mathcal{C}^{\textrm{gain}\downarrow}\bar{\rho}^{\uparrow}(E,r).
\end{eqnarray}
\end{subequations}
Our aim is to find a solution to these equations 
subject to the boundary conditions  
\begin{eqnarray}
\label{bc1}
\rho^{\uparrow ee}(E,R_{\nu}) &  =  &  \kappa^{e} \   f^{e}(E)\\
\label{bc2}
\rho^{\uparrow xx}(E,R_{\nu}) & = & \kappa^{x} \     f^{x}(E)\\
\label{bc3}
\bar{\rho}^{\uparrow ee}(E,R_{\nu}) & = & \kappa^{\bar{e}}  \  f^{\bar{e}}(E)\\
\label{bc4}
\bar{\rho}^{\uparrow xx}(E,R_{\nu}) & =  &  \kappa^{x} f^{x}(E)\\
\label{bc5}
\rho^{\uparrow ex}(E,R_{\nu}) &  =  &  0\\
\label{bco1}
\bar{\rho}^{\uparrow ex}(E,R_{\nu}) & = & 0\\
\label{bco2}
\rho^{\downarrow}(E,r^{\textrm{max}}) &=& 0 \\
\label{bc6}
\bar{\rho}^{\downarrow}(E,r^{\textrm{max}}) &=& 0~,
\end{eqnarray}
where $r^{\textrm{max}} \to  \infty$. 
Here $f^{i}$ are the initial thermal spectra for various flavors at the neutrino sphere given by the Fermi-Dirac distribution,
\begin{eqnarray}
f^{i}(E) = \frac{E^{2}}{1+\exp(E/T_{i}-\eta_{i})}.
\end{eqnarray}
The normalization constants, 
\begin{eqnarray}
\kappa^{i}=\frac{1}{\int_{0}^{\infty}f^{i}(E)dE}\frac{j_{i}}{j_{e}}, 
\end{eqnarray}
are used to normalize the thermal spectra, which sets the scale for $\rho_{ee}$ in Figs.~\ref{onset}-\ref{flux}. 
Here, $j_{i}$ is the flux of the $i^{\textrm{th}}$ flavor of the neutrinos which is equal to the ratio of luminosity and the average energy, $\frac{L_{i}}{\langle E_{i} \rangle}$.
The flux of neutrinos, which is the coefficient of the self-interaction Hamiltonian, is adjusted to have luminosity of $10^{51}$ ergs/sec for all flavors. We also assume that the degeneracy factors
$\eta_{i} = \mu_i/T$ are the same for all flavors and  are  equal to $3.0$. We use temperature values of $T_{e}=2.76$ MeV, $T_{\bar{e}} = 4.01$ MeV and $T_{x} = T_{\bar{x}} = 6.26$ MeV \cite{Keil:2002in}.

We now  describe in detail each term on the RHS of eqs.~(\ref{eq:qkes}).  
First, note that the factor of $ 1/ \cos\vartheta$ on the right hand side of eq.~\eqref{eom1},\eqref{eom2},\eqref{eom3},\eqref{eom4} 
is a geometrical factor to convert path length traversed by the neutrino to radius.

Next, the  four  Hamiltonians  determining the coherent evolution of density matrices  are given by,
\begin{eqnarray}
H^{\uparrow(\downarrow)} &=& H_{0}+H_{\textrm{m}}+H_{\nu\nu}^{\uparrow(\downarrow)}\\
\bar{H}^{\uparrow(\downarrow)} &=& -H_{0}+H_{\textrm{m}}+H_{\nu\nu}^{\uparrow(\downarrow)}
\end{eqnarray}
where,
\begin{eqnarray}
H_{0} &=& \frac{1}{2}
\begin{pmatrix}
-\omega \cos 2 \theta_{\textrm{V}} & \omega \sin 2 \theta_{\textrm{V}} \\
\omega \sin 2 \theta_{\textrm{V}} & \omega \cos 2 \theta_{\textrm{V}}
\end{pmatrix}\\
H_{\textrm{m}} &=&
\begin{pmatrix}
\sqrt{2} G_{F} n_{e} & 0 \\
0 & 0
\end{pmatrix}\\ 
H^{\uparrow}_{\nu\nu} &=&  \mu  \  \int_{0}^{\infty}dE'~  \left(\rho^{\uparrow} (E',r)  -\bar{\rho}^{\uparrow} (E',r) \right) (1-\cos^{2}\vartheta) \nonumber\\
       & + &\mu  \int_{0}^{\infty}dE'~  \left( \rho^{\downarrow} (E',r) -\bar{\rho}^{\downarrow}  (E',r)  \right)(1+\cos^{2}\vartheta),\\
H_{\nu\nu}^{\downarrow} &=&   \mu \ \int_{0}^{\infty}dE'~  \left(\rho^{\downarrow}  (E',r)-\bar{\rho}^{\downarrow} (E',r)\right) (1-\cos^{2}\vartheta) \nonumber\\
        &+& \mu \   \int_{0}^{\infty}dE'~  \left( \rho^{\uparrow} (E',r)-\bar{\rho}^{\uparrow}  (E',r) \right)(1+\cos^{2}\vartheta).
\end{eqnarray}

$\omega$ is the vacuum oscillation frequency which is equal to $\frac{\Delta m^{2}}{2E}$. The coefficient of self-interaction potential, $\mu$,  is adjusted so that the total neutrino luminosity for all the flavors is $10^{51}$ ergs/sec, 
and takes the form
  $\mu (r) = \sqrt{2} G_F  n_\nu (r) $ with 
$n_\nu (r) = \frac{L_{\nu_{e}}}{\langle E_{\nu_{e}}  \rangle 2 \pi r^{2}}\frac{\cos\vartheta_{0}}{\cos\vartheta}$.

Finally,  $\mathcal{C}^{\uparrow} (E)$ and $\mathcal{C}^{\downarrow} (E)$ are the collision terms for outgoing and incoming  neutrinos beams. 
There are two  terms for each bin,  corresponding to  loss and gain (i.e. scattering out of or into the given bin), as indicated by the superscript.
The form of the collision term used in eqs.~\eqref{eom1} can be derived from the general results of Ref.~\cite{Blaschke:2016xxt}   
 for neutrino-nucleon (or neutrino-nucleus) scattering,   
under the following assumptions:
(i)  neglecting Pauli blocking for both neutrinos and targets: this corresponds to non-degenerate species; 
(ii)  treating targets as  very heavy and non-relativistic ($M \gg T$): this implies that collisions do not change the neutrino energy, 
which is an excellent approximation for the halo problem of neutrino scattering off nuclei; 
(iii) performing the angular integrals consistently with the single-angle approximation:  
this implies that  a neutrino of momentum $(|\vec{k}|, \cos\vartheta)$ can only scatter into a final state with 
momentum $(|\vec{k|},  \pm \cos\vartheta)$ (forward or backwards cones). 

The individual terms have to satisfy  $\mathcal{C}^{\textrm{loss}\uparrow}= \mathcal{C}^{\textrm{gain}\downarrow}$ 
and  $\mathcal{C}^{\textrm{loss}\downarrow}= \mathcal{C}^{\textrm{gain}\uparrow}$ to be consistent with the principle of detailed balance.
Explicit calculation as sketched above reveals that $\mathcal{C}^{\textrm{loss}\uparrow}= \mathcal{C}^{\textrm{gain}\uparrow}$, 
namely all entries are equal. 
The explicit calculation also shows that $\mathcal{C}^{\textrm{loss}\uparrow} = n_T   G_F^2 E^2  C_W /\pi$, where 
$n_T$ is the number density of targets,  $G_F$ is the Fermi constant, and  $C_W$ 
is a  coupling of $O(1)$ proportional to  the weak charge of the target.

An order-of-magnitude estimate of   $\mathcal{C}^{\textrm{loss}\uparrow}$ can be obtained as follows. 
First, note that for neutrino-nucleon scattering the collision term goes like $\sim G_{F}^{2} E^{2} n_{e}$ while the matter potential goes like $\sim G_{F} n_{e}$. 
The collision term is thus related to the matter potential by a factor of about $\sim G_{F} \langle E^{2} \rangle \approx 10^{-9}$. Notice that by taking the average of the energy-squared we have ignored the energy dependence which does not affect the results significantly. 
Since the cross section can be enhanced by two or three orders of magnitude for scattering off  nuclei (through the coherence factor of  $A^2$), 
 it is quite reasonable   to estimate the maximum effect of the collision term by using $\mathcal{C}\sim 10^{-6} \sqrt{2} G_{F} n_{e}$. We use a matter profile that falls off with the third power of the radius with an entropy per baryon of 140. Choosing numerical parameters this way corresponds to about a percent of the neutrinos experiencing direction-changing scatterings, compared to an estimate of 0.1 percent mentioned in ~\cite{Cherry:2012zw}. We have purposely used a collision term which is almost an order of magnitude larger.  

\section{Numerical results}
\label{sec:nr}
%**********************************************************************
It should be noted that the boundary conditions for eq.~\eqref{bc1},~\eqref{bc2},~\eqref{bc3},~\eqref{bc4} and eq. \eqref{bc5}, \eqref{bc6} are at two different radii. This makes the problem unsolvable by a direct application of the forward-difference method. we employ an iterative forward-difference method to solve this boundary value problem.

In this section we describe the methodology we use and the numerical results we obtain. For the zeroth iteration we solve the equations of motion and evolve $\rho$ and $\bar{\rho}$ from $r^{\textrm{min}}$ to $r^{\textrm{max}}$, ignoring the Halo effect ($\mathcal{C}_{\textrm{loss}}=\mathcal{C}_{\textrm{gain}}=0$). Then using eq.~\eqref{eom3},~\eqref{eom4} we evolve $\rho^{\downarrow}$ and $\bar{\rho}^{\downarrow}$ from $r^{\textrm{max}}$ to $r^{\textrm{min}}$ using $\rho$ and $\bar{\rho}$ from the previous step. Now that we have an initial estimate for $\rho^{\downarrow}$ and $\bar{\rho}^{\downarrow}$ we use those to solve eq.~\eqref{eom1} and \eqref{eom2} and proceed with the next iterative steps. For our computation we use $r_{\textrm{min}}=15$ km and $r_{\textrm{max}} = 515$ km.

For all but the first iteration in the radially outward direction, values of $\rho^{\downarrow}$ are required for calculation of $\rho^{\uparrow}$ and vice-versa. The values required are not on a fixed grid but are determined by adaptive Runge-Kutta. After each calculation of $\rho^{\uparrow}$, we store the radial dependence of each energy bin as a spline curve for use in the following calculation of $\rho^{\downarrow}$ and then store those density matrices as spline curves for use in the following calculation of  $\rho^{\uparrow}$. We use 80 energy bins in the range of 0 to 80 MeV and there are 8 components of density matrix for each energy bin. We thus store 640 spline curves in memory at all times during the evolution of the code. We use the publicly available GNU Scientific Library (GSL) to perform this task~\cite{Gough:2009:GSL:1538674}. 

\begin{figure*}
\includegraphics[width=0.49\textwidth]{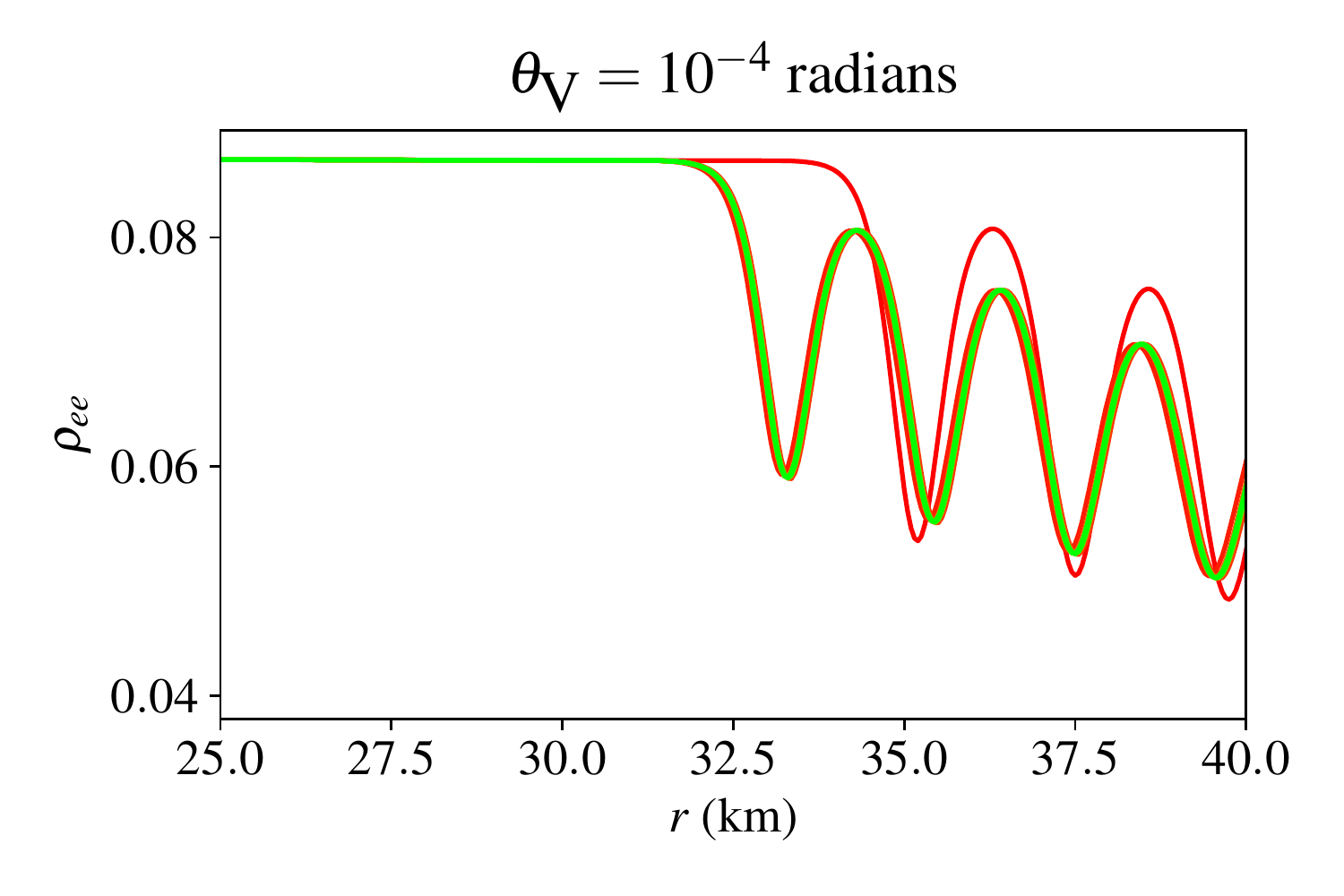}
\includegraphics[width=0.49\textwidth]{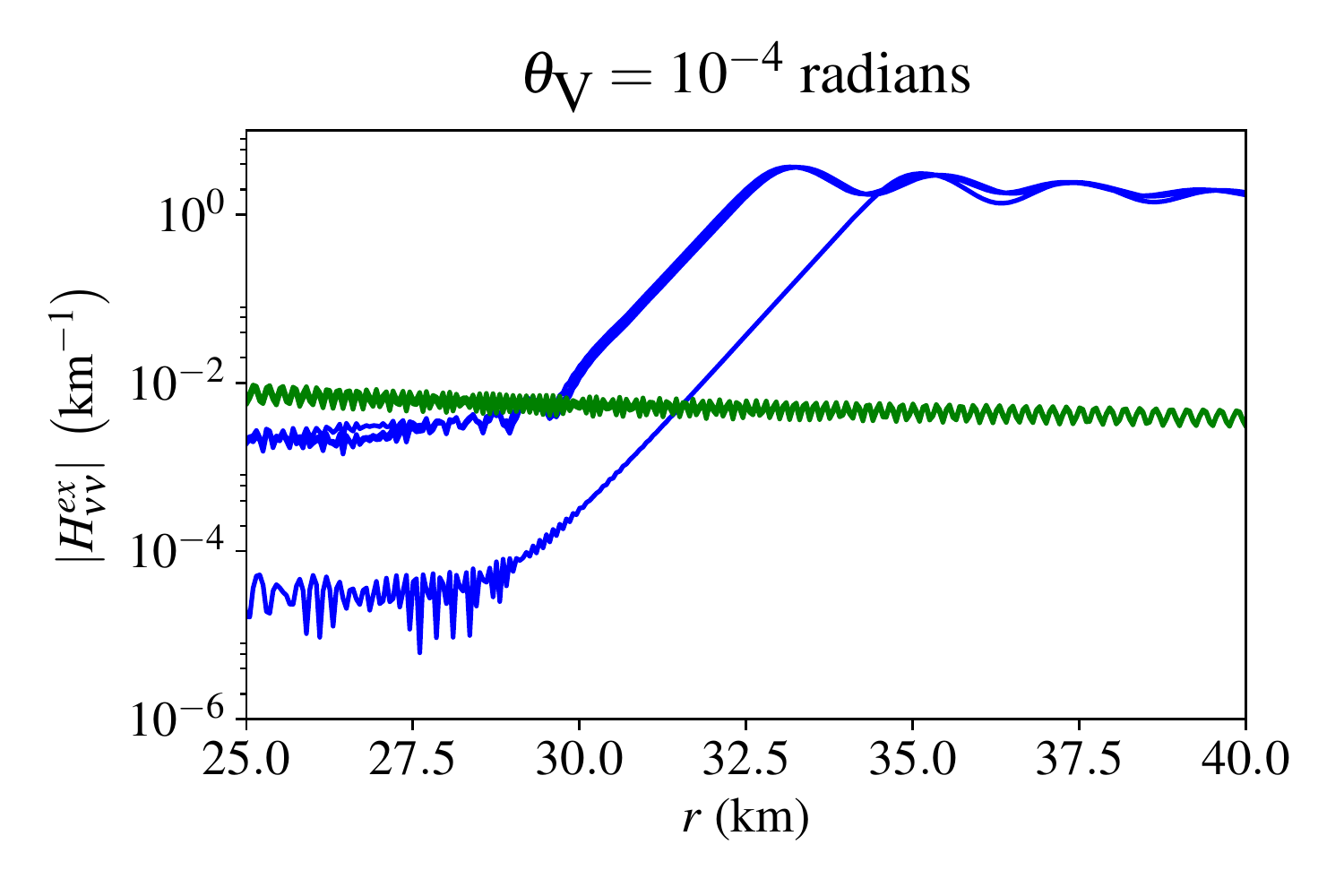}
\caption{The onset of neutrino oscillations on the left for energy bin of 8.5 MeV. The left plot has 15 iterations which gradually go from red to green, however only two plots are visible as the effect of Halo converges after the second iteration. On the right we plot the off-diagonal term of the Hamiltonian on log scale in blue for 15 iterations. The blue one which starts off with small magnitude is the first iteration. On the right hand side we also plot the off-diagonal term of the Hamiltonian due to the Halo term only in green. The vacuum mixing angle is set to $10^{-4}$ radians.}
\label{th1e-4}
\end{figure*}

\begin{figure*}
\includegraphics[width=0.49\textwidth]{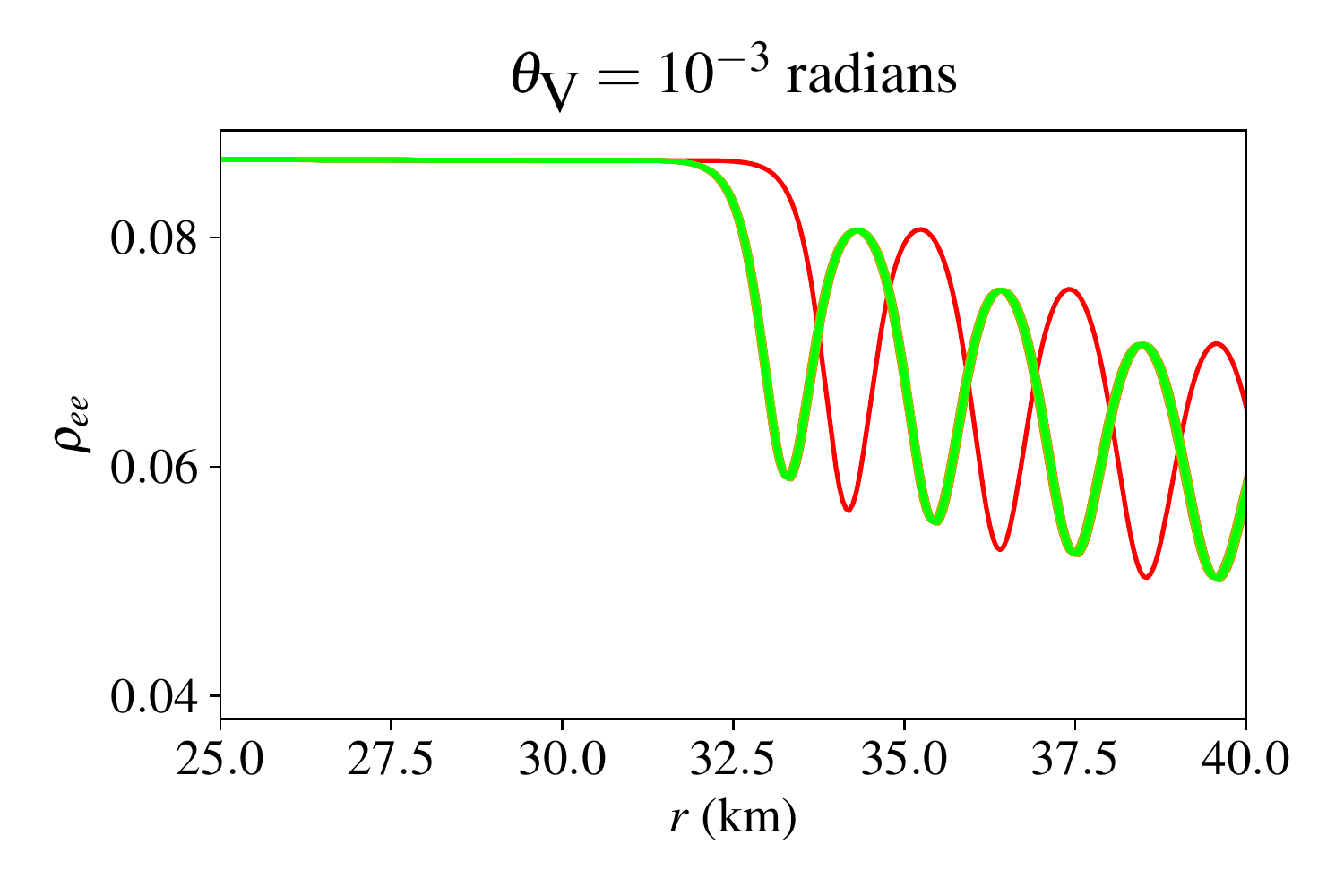}
\includegraphics[width=0.49\textwidth]{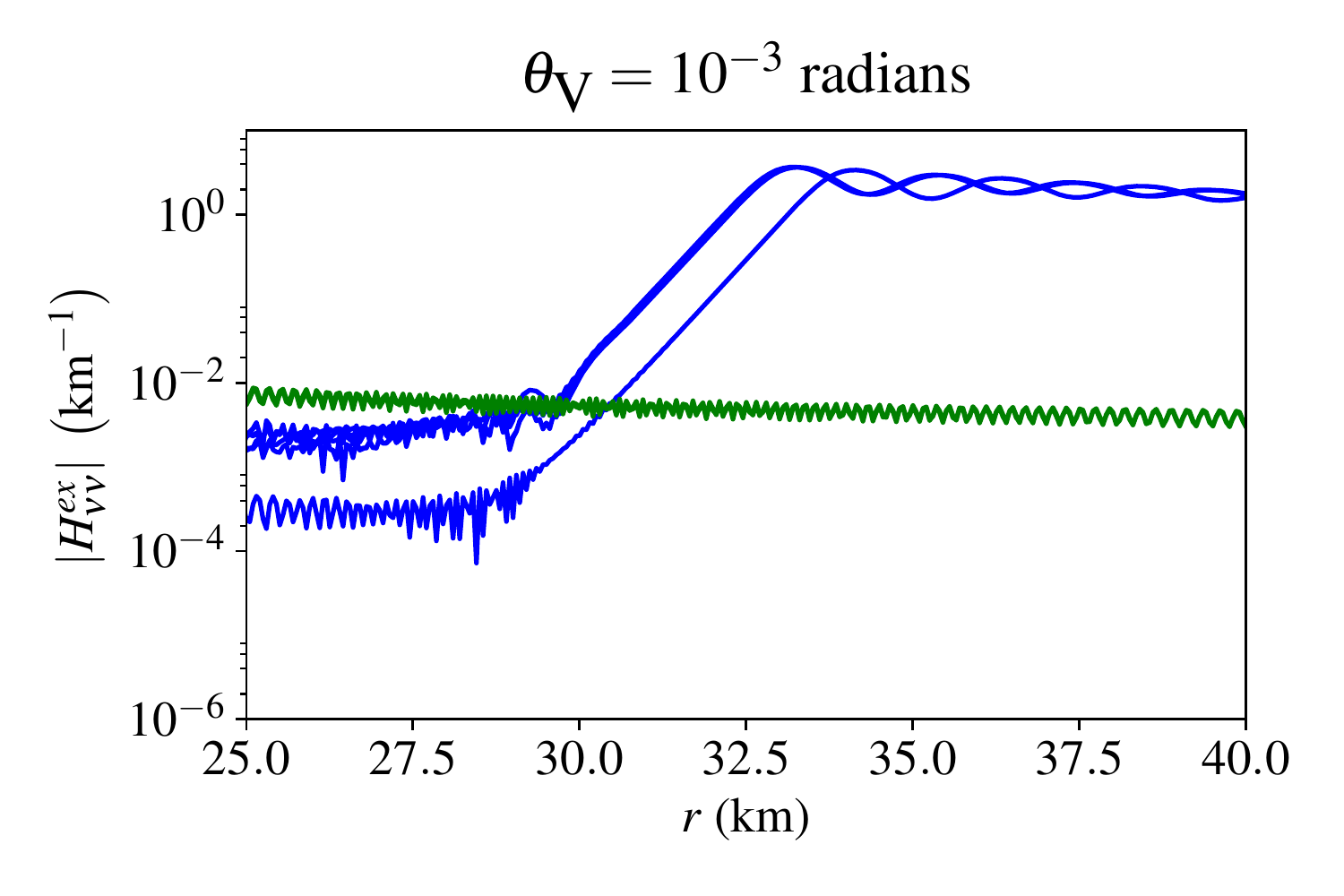}
\caption{Same as Fig.~\ref{th1e-4} but for vacuum mixing angle set to $10^{-3}$ radians.}
\label{th1e-3}
\end{figure*}

\begin{figure*}
\includegraphics[width=0.49\textwidth]{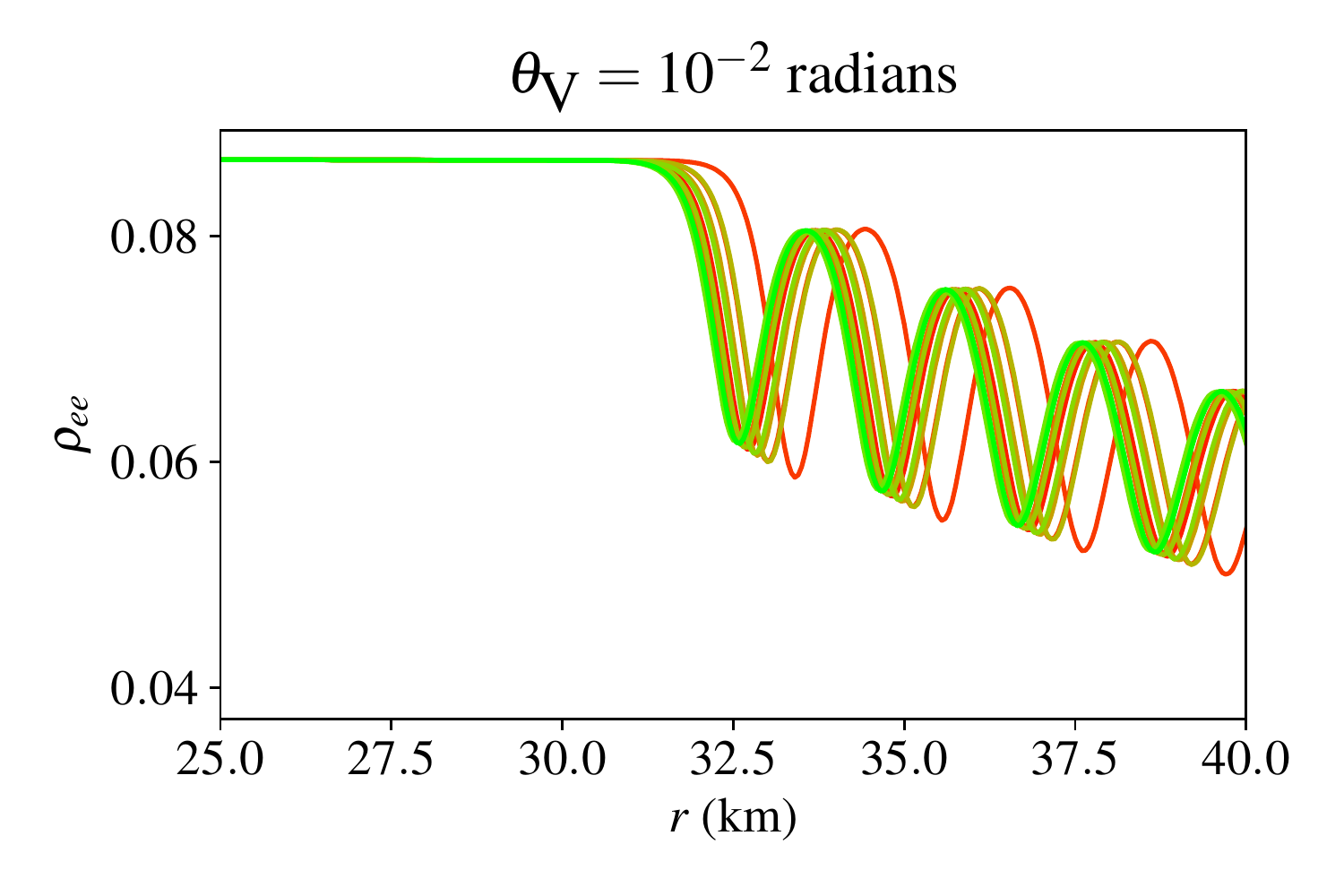}
\includegraphics[width=0.49\textwidth]{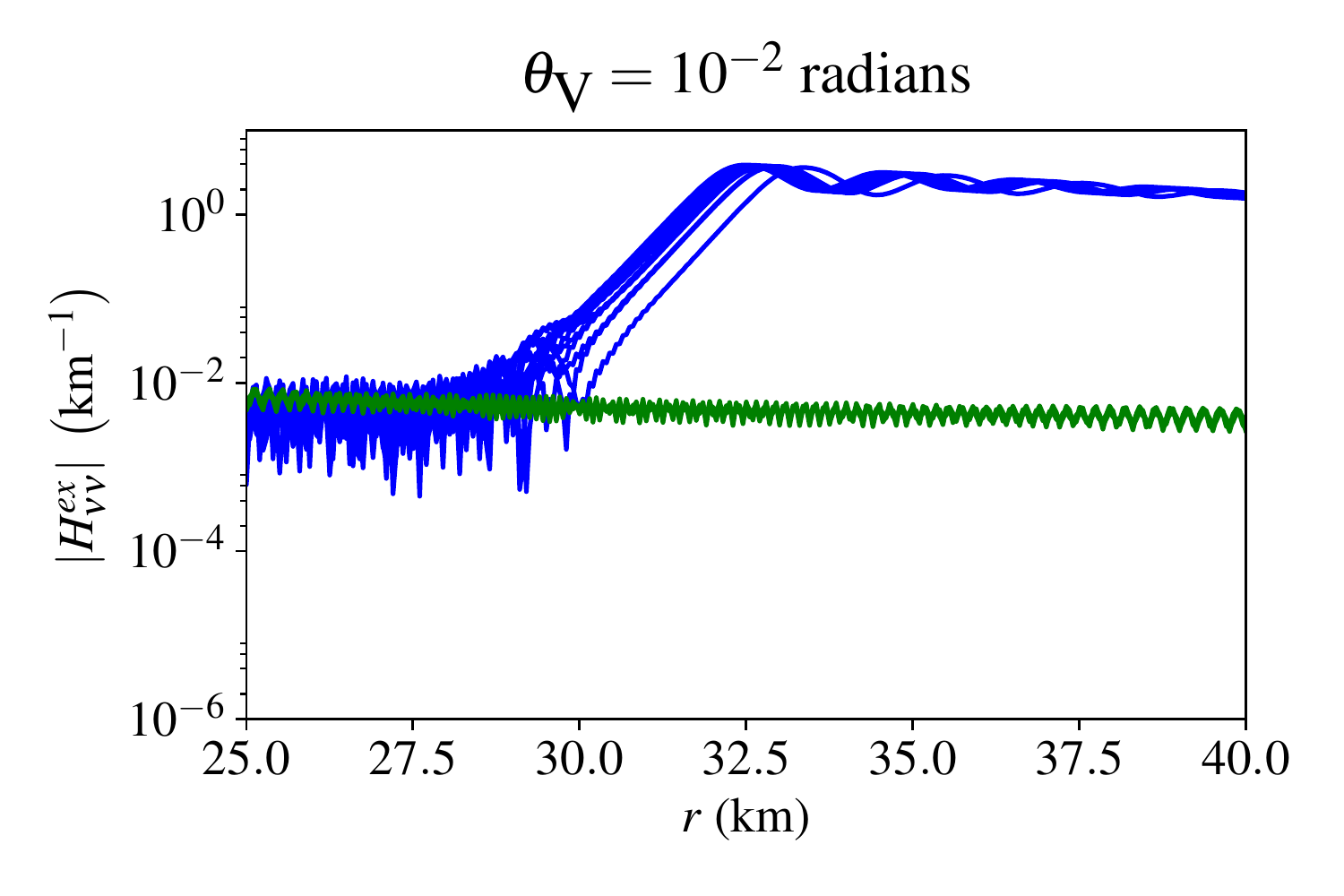}
\caption{Same as Fig.~\ref{th1e-4} but for vacuum mixing angle set to $10^{-2}$ radians.}
\label{th1e-2}
\end{figure*}

On the left panels of Figs.~\ref{th1e-4}, \ref{th1e-3} and \ref{th1e-2} we plot $\rho_{ee}$ for the 10th energy bin corresponding to the energy of 8.5 MeV for 15 iterations in the region of neutrino flavor onset. On the right panels, we plot the off-diagonal term of the self-interaction Hamiltonian for 15 different iterations and the off-diagonal for the contribution due to the Halo effect alone.
Figs.~\ref{th1e-4}, \ref{th1e-3} and \ref{th1e-2} correspond to different  values for the vacuum mixing angle,  namely  $\theta_V = 10^{-4}, 10^{-3}, 10^{-2}$.  
 It can be clearly seen that the for small values of the mixing angle, when initially the contribution of the Halo effect is large compared to the off-diagonal term without the Halo effect, the convergence is swift and happens mostly by the second iteration. On the other hand, when the contribution of the Halo term to the off-diagonal term is comparable to the off-diagonal term without the Halo effect, the convergence requires a few iterations, but the overall effect remains small. 

It should be noted  that the off-diagonal term of $H_{\nu \nu}$  without the Halo effect is proportional to $(1-\cos^{2}\vartheta)$ which in turn depends on $\vartheta_{0}$, the emission angle which is fixed in our case to $5^{\circ}$. For larger values of $\vartheta_{0}$, the off-diagonal term is larger to begin with and the Halo's impact is less pronounced. For smaller emission angles the relative effect of the Halo is larger but the flux from small emission angle is suppressed due to the geometric factor, eq.~\ref{varthetadef}. 

From the right panels of Figs. \ref{th1e-4}, \ref{th1e-3} and \ref{th1e-2} we can see that the back-scattering increases the  small $r$  value of the off-diagonal element of the
self-interaction  Hamiltonian $H_{\nu \nu}$. The radius at which the exponential growth of the off-diagonal element begins and the value of the exponent remain unchanged, and the two effects together lead to an onset of flavor transformation at a smaller radius, reduced by a few kilometers with respect to the standard bulb model. 

%\onecolumngrid

%\twocolumngrid

\section{Conclusion}
\label{sec:conc}
%**********************************************************************

\begin{figure}
\begin{center}
\includegraphics[width=0.49\textwidth]{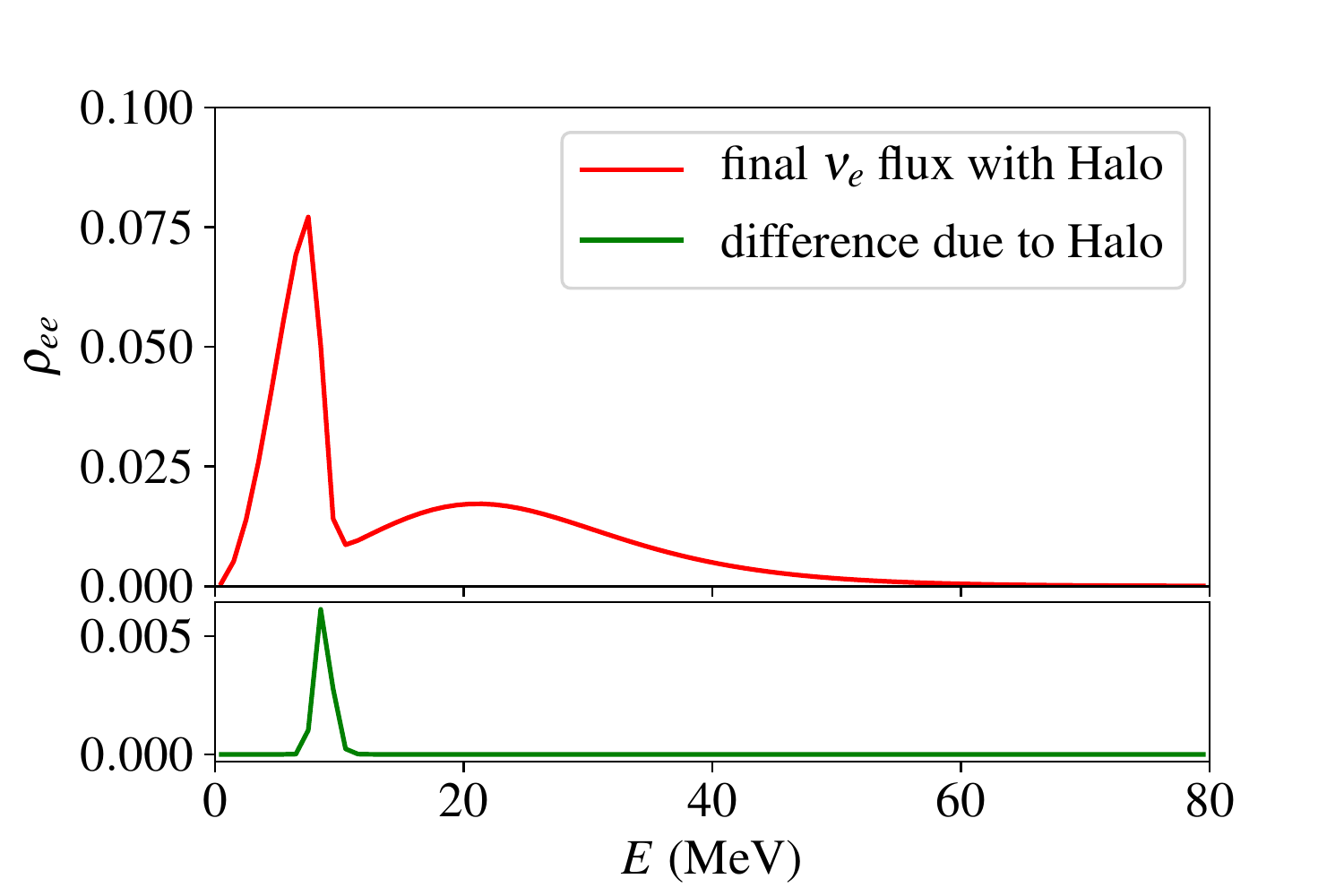}
\includegraphics[width=0.49\textwidth]{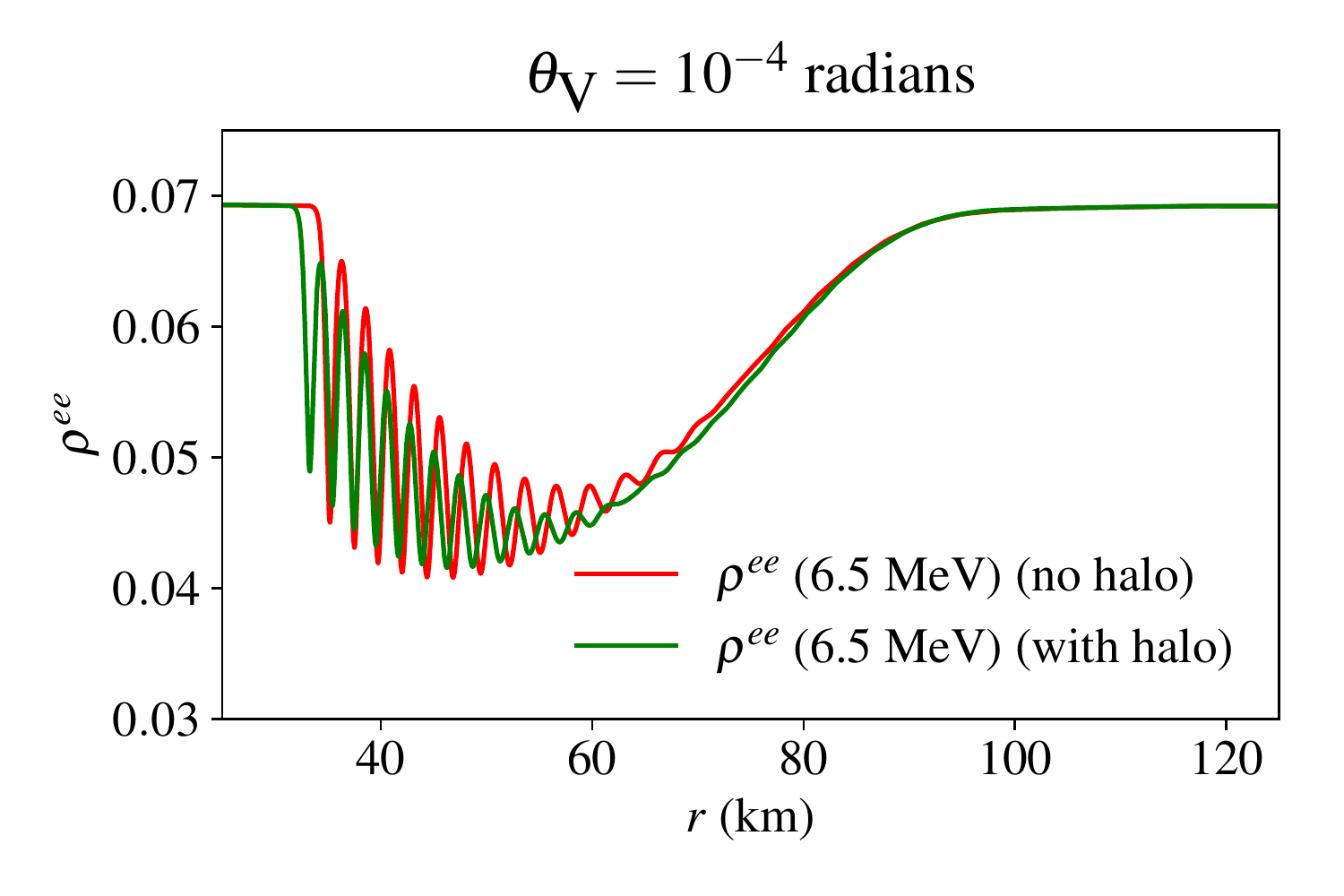}
\caption{Final electron neutrino flux with the Halo effect. In the left panel we can see that the `spectral split' feature; the discontinuity in the flux is preserved when Halo effect is included. For comparison we also plot the difference between the results with no Halo effect and the one with Halo effect. The Halo effect has a very limited influence on the shape of the final neutrino spectra. In the right hand side panel we plot the flux as a function of radius for a single energy. We can see that although the radius of flavor onset is changed by the inclusion of the Halo effect, the final flux remains unchanged. }
\label{flux}
\end{center}
\end{figure}

In this paper  we have formulated the problem of collective neutrino oscillations 
with the Halo effect in terms of appropriate QKEs with simplified collision terms,  and have adapted those to the single-angle approximation. 
We have then obtained a self-consistent solution for the resulting boundary value problem using an iterative method.
Our main findings can be summarized as follows:

\begin{itemize}

\item  Halo effects modify the size of the various components of the Hamiltonian $H_{\nu \nu}$   consistently with the expectations of Ref.~\cite{Cherry:2012zw}.
In particular, the modification can be dramatic at large radii.  At this large distance where the neutrino flavor field has stopped evolving, the Halo potential is large compared to the self-interaction potential, but it does not trigger additional flavor instabilities.  
On the other hand the influence of the Halo potential is more important in the region near the onset of neutrino flavor instability.

\item   At the onset of neutrino flavor instability,  the main impact of the Halo is to   increase the 
 value of the off-diagonal element of the   self-interaction  Hamiltonian $H_{\nu \nu}$ by about two orders of magnitude (for small vacuum mixing angle).  
On the other hand,  at these radii the Halo affects the diagonal elements of $H_{\nu \nu}$ at the sub-percent level, 
and hence it has little impact on the radius at which the instability arises and on the rate of exponential growth of the off-diagonal elements of $H_{\nu \nu}$. 
These features  together lead to an onset of flavor transformation at a smaller radius, reduced by a few kilometers, as illustrated in Figs.~\ref{th1e-4}, \ref{th1e-3} and \ref{th1e-2}.

\item An important question that has been repeatedly raised in the literature regarding the collective neutrino oscillations in the vicinity of core-collapse supernova is about the robustness of the split. At least in the case of single angle approximation, the feature of split seems to survive the inclusion of the Halo effect, as illustrated in Fig.~\ref{flux}.

\item We find that in the simplest of the supernova models, assuming spherical symmetry, the Halo  has an effect which is equivalent to increasing the effective mixing angle -- a conclusion we expect would carry on to the multi-angle calculations. The overall impact of the Halo effect is thus expected to be very limited in a spherical model of neutrino emission from a supernova.

\end{itemize}

Looking ahead,  our work can be extended and generalized in several directions, with different degree of  technical difficulty.
First,   let us note that the matter profile we use, and hence the radial dependence of the strength of the Halo effect,  is smooth. However, in the interior of a supernova environment, there can be sharp changes in matter density and the nuclear composition. The effect of an abrupt change in the strength of the collisions certainly warrants further investigation. 
Finally,   while it seems quite plausible that the qualitative effects of the Halo  found in this single angle calculation  
survive in a realistic model, we think that  future work should be devoted to exploring a multi-angle scheme.

%**********************************************************************   
\section{Acknowledgments}
We are extremely grateful to George Fuller and Baha Balantekin for many engaging discussions. 
SS would like to thank the N3AS collaboration for the hospitality and opportunity to discuss this work with several colleagues during the first annual N3AS collaboration meeting in University of California, San Diego. 
Research presented in this article was supported by the Laboratory Directed Research and Development program of Los Alamos National Laboratory under Project \#20170430ER.

%**********************************************************************
%**********************************************************************
%**********************************************************************   

%\bibliographystyle{apsrev4-1}
\bibliographystyle{JHEP}
\bibliography{halo3}

\end{document}